\documentclass[a4paper,12pt]{article}
\usepackage{graphicx,color,pifont,psfrag,bm}
\usepackage[dvipsnames]{xcolor}
\usepackage{amsmath,amssymb,a4wide,epsfig,mathtools}
\usepackage{soul}
\usepackage[makeroom]{cancel}
\usepackage{lineno}
\usepackage{cite}
\usepackage[bookmarks=false,colorlinks]{hyperref}
\hypersetup{urlcolor=blue, citecolor=blue} 
\date{\today}
 \normalsize

\newcommand{\bmat}{\left(\begin{array}}
\newcommand{\emat}{\end{array}\right)}
\newcommand{\be}{\begin{equation}}
\newcommand{\ee}{\end{equation}}
\newcommand{\bea}{\begin{eqnarray}}
\newcommand{\eea}{\end{eqnarray}}

\def\Bbar{\overline{B}}

\def\jay{\mathcal{J}}
\def\see{\mathcal{C}}
\def\lsim{\raise0.3ex\hbox{$\;<$\kern-0.75em\raise-1.1ex\hbox{$\sim\;$}}}
\def\gsim{\raise0.3ex\hbox{$\;>$\kern-0.75em\raise-1.1ex\hbox{$\sim\;$}}}

\definecolor{grey}{rgb}{0.3,0.3,0.3}

\begin{document}
\bibliographystyle{unsrt}
\renewcommand{\thefootnote}{\fnsymbol{footnote}}
\rightline{{LPT-Orsay-16-29}} \rightline{LAL-16-010}
\vspace{.3cm} 
{\Large
\begin{center}
{\bf Angular analysis of $\bm{B\to J/\psi K_1}$ : towards a model independent determination of 
the photon polarization with $\bm{B\to K_1\gamma}$ }
\end{center}}
\vspace{.3cm}

\begin{center}
E. Kou$^a$, A. Le Yaouanc$^b$, A. Tayduganov$^{c}$\\
\vspace{.3cm}\small
\emph{$^a$ 
Laboratoire de l'Acc\'el\'erateur Lin\'eaire,
  Univ. Paris-Sud, CNRS/IN2P3 (UMR 8607)} \\
\emph{Universit\'e Paris-Saclay, 91898 Orsay  C\'{e}dex, France}

\emph{$^b$ Laboratoire de Physique Th\'eorique, 
CNRS/Univ. Paris-Sud 11 (UMR 8627)}\\
\emph{91405 Orsay, France}

\emph{$^c$ 
{CPPM, Aix-Marseille Universit\'{e}, CNRS/IN2P3 and Aix Marseille Universit\'{e},}} \\
\emph{{Universit\'{e} de Toulon, CNRS, CPT UMR 7332, 13288, Marseille, France}} 
\\ \vskip 0.5cm
\today
\end{center}
\vskip .3cm
%Version EK ALY March 2016 v1 modified by Alain in march\\
%Version EK ALY Fev 2016 v1 modified by Emi in fev.\\
\hrule \vskip 0.3cm
\begin{center}
\small{\bf Abstract}\\[3mm]
\end{center}
We propose a model independent extraction of the hadronic information needed to determine the photon polarization of the $b\to s\gamma$ process by the method utilizing the $B\to K_1\gamma \to K\pi\pi {\gamma}$ angular distribution. We show that exactly the same hadronic information can be obtained by using the $B\to J/\psi K_1 \to J/\psi K\pi\pi$ channel, which leads to {a} much higher precision.
\begin{minipage}[h]{14.0cm}
\end{minipage}
\vskip 0.3cm \hrule \vskip 1.2cm
%\newpage
%%%%%%%%%%%%%%%%%%%%%%%%%%%%%%%%%%
%

%\linenumbers

\section{Introduction}\label{sec:1}
The circular polarization of the photon in the $b\to s\gamma$ process has {a} unique sensitivity to new physics, namely to the right-handed charged current (see e.g.~\cite{Becirevic:2012dx,Kou:2013gna,Haba:2015gwa}). While it is a very fundamental observable, the experimental determination of the photon polarization was not achieved at a high precision in the previous $B$ factory experiments. Therefore, this is a very important challenge for LHCb as well as for the upgrade of $B$ factory, Belle~II experiment. Various theoretical ideas to measure the photon polarization have been proposed {(pioneered  by~\cite{Atwood:1997zr,Atwood:2007qh,Gronau:2001ng,Gronau:2002rz} and followed by~\cite{Kou:2010kn, Bishara:2015yta, Muheim:2008vu,Oliver:2010im}}) and many experimental efforts are currently on-going~\cite{exp}. {Since the photon polarization measurement determine the Wilson coefficient $C_7^{(\prime)}$, it will have an important consequence to the global fit as well~\cite{theory}.}

Recently the LHCb collaboration has presented an interesting result~\cite{Aaij:2014wgo} on the so-called up-down asymmetry of the $B\to K\pi\pi\gamma$ decay, originally proposed in~\cite{Gronau:2001ng,Gronau:2002rz}. The up-down asymmetry, which is the difference of the number of events with photon {emitted} above and below the $K\pi\pi$ decay plane {in the $K\pi\pi$ reference frame}, can indeed provide the information {on} the photon polarization. 
The basic idea is to determine the photon polarization by measuring the $K_1$ polarization, 
which is correlated {with} the photon polarization, through its angular distribution 
in the {$B\to K\pi\pi\gamma$} decay. 

{To determine the photon polarization from the LHCb result}, we need the detailed prediction of the $K_1\to K\pi\pi$ strong decay. In our previous works~\cite{Kou:2010kn,Tayduganov:2011ui}, we have obtained this information by using the other experimental results, mainly the isobar model description from the ACCMOR collaboration~\cite{Daum:1981hb}, complemented by the theoretical model computation using the $^3P_0$ model~\cite{LeYaouanc:1972vsx}. The $B\to K_1(1270)\gamma\to K\pi\pi\gamma$ channel, different from {the} $K_1(1400)$ channel,  requires various unconventional treatments and unfortunately, our conclusion is that there are certain {uncertainties} remaining to describe this channel. The main difficulties are (see~\cite{Tayduganov:2011ui} for the detailed discussions) : 
\begin{itemize}
\item {the existence of two intermediate processes, $K_1(1270)\to K^* \pi$ and $K_1(1270)\to K \rho$, with the latter being just on the edge of the $K \rho$ phase space and having however a large branching ratio. Quasi-threshold effects must be taken into account.}

\item {furthermore, as  we found, the final estimation of photon polarization is also sensitive to the {contribution of the} $K_1(1270)$ decay channels with scalar isobars,  $K_1(1270) \to K (\pi \pi)_{S-wave}$ or $K_1(1270) \to (K \pi)_{S-wave} \pi$, which are not well determined, neither by experiment nor by theory.}
\end{itemize}
These problems must be solved in the future with more detailed analysis of $K_1$ resonances, which are produced from $B$, $\tau$ or $J/\psi$ decays. 

In this article, we rather propose a model independent approach to circumvent the problem. {In all the previous works, only a partial angular distribution was considered, i.e. taking into account only one $\theta$ angle. We show in this article that with a more complete angular description, the information on the $K_1$ decay  needed for photon polarization determination can be extracted  directly from $B \to K \pi \pi + \gamma$  decay. That is, using the angles involving not only the $\cos\theta$ like distribution which yields the up-down asymmetry, but also the azimuthal angle $\phi$ dependence, we can obtain the full hadronic information without the isobar model description of the resonances. }

In fact, with the limited statistics available for $B \to K \pi \pi + \gamma$, this method is {currently difficult}. On the other hand, it turns out that we can obtain the same hadronic information from another channel $B \to K \pi \pi + J/\psi$ where two orders of magnitudes higher statistics{, with respect to the photon channel,} {is} available~\cite{Guler:2010if}. We show that the full angular distribution measurement allows us to separate the $B$ decay and $K_1$ decay part{s} so that we can extract {the} same hadronic information from the $B\to K \pi\pi + J/\psi$ decay. 

For the moment, for a simpler illustration of the approach, we consider the case of only one $K_1$ resonance, which may be practically supported by the the fact that $B \to K_1(1270) \gamma$ seems largely dominant over $B \to K_1(1400) \gamma$~\cite{Yang:2004as}.

The {rest} of the article is organized as follows: in section 2, we introduce the decay amplitudes of {
$B \to K_1 J/\psi$ and $B \to K_1\gamma $ with $K_1$ decaying to $K\pi\pi$}. In section 3, we derive the angular distribution{s} for {these decays}. Then, we demonstrate  in section 4 that the hadronic information we need to determine the photon polarization {in}  $B \to K_1 \gamma $ can be obtained directly from {the measurement of angular coefficients in} $B \to K_1 J/\psi$ and/or $B\to K_1\gamma$, and we conclude in section 5. 

%%%%%%%%%%%%%%%%%%%%%%%%%%%%%%%%%%%%%%%%%%%%%
\section{The decay amplitudes and rates}
The four body decay rate can be written as the product of the decay {rates} of $B \to K_{1s_z} V_{s_z}$ and $K_{1{s_z}}\to K\pi\pi$  summed over the different $V$ polarization{s}
%, and times the squared propagator of the
\footnote{We follow the PDG convention, i.e. $
\int_{\Omega} d\Phi_2 = \frac{1}{(2\pi)^5}\frac{|\vec{p}_V^{{\,*}}|}{{2}M_B}, \ 
\int_{\psi} d\Phi_3 = \frac{1}{32(2\pi)^8}\frac{1}{s}ds_{13}ds_{23}d\phi d(\cos\theta)$.} : 
\bea
d\Gamma_4^V(s)&\equiv& d\Gamma(B\to K_{1} V\to (K\pi\pi)V)_s  \label{eq:1}\\
&=& \sum_{s_z}
\frac{(2\pi)^4}{2M_B}\Big|
%\sum_{K_1=K_1(1270), K_1(1400)}
{\mathcal{M}}_{s_z}^{{V}}(B\to K_{1s_z}V \to (K\pi\pi)V)_s\Big|^2 (2\pi)^3
%\sqrt{s}
{ds} d\Phi_2d\Phi_3 \,, \nonumber 
\eea
where $s_z$ is the polarization of $V=J/\psi, \gamma$ : 
\be
s_z=0, \pm 1 \quad {\rm (for\ }V=J/\psi), \quad \quad  
s_z=\pm 1 \quad {\rm (for\ }V=\gamma)  \,.
\ee
{Here,  $B$  can be $B^\pm$, $B^{{0}}$ or $\Bbar^{{0}}$. }
{Denoting the amplitude of 
$B\to K_{1}(s) V$ as $\mathcal{A}_{s_z}(s)$ and of $K_{1}(s)\to K\pi\pi$  as $\epsilon_{K_{1s_z}}^\mu\jay_\mu$ \,, one can write} : 
\be
\mathcal{M}_{s_z}^{{V}}(B\to K_{1s_z}V \to (K\pi\pi)V)_s =\frac { \mathcal{A}_{s_z}^V(s) \times (\epsilon_{K_{1s_z}}^{\mu} \jay_\mu (s_{13},s_{23})_s)} {(s-m_{K_1}^2)+i m_{K_1} \Gamma_{K_1}(s)} \,.
\label{eq:6}
\ee
In the following, we consider only $K_1=K_1(1270)$ for simplicity, {though} it can be readily extended to include $K_1(1400)$.  The propagator of the $K_1$, which is parametrized here as Breit-Wigner function, is introduced in order to use the $K \pi \pi$ invariant mass {$m_{K\pi\pi}\equiv\sqrt{s}$ as the varying $K_1$ mass}. {{The $K_1$ rest frame is meant as the actual $K \pi \pi$ system.}} This is not a convention, but an assumption on the off-shell extrapolation of amplitudes, partially justified by unitarity. Note that this impl{ies} that the Dalitz plot $(s_{13},s_{23})$ depends on $s$ as well. 

In Eq.~\eqref{eq:6}, the full kinematical variable dependence of $\jay$ is left implicit but it can be displayed with help of two form factors as $\see_{1,2}$~\cite{Kou:2010kn}: 
\be
\jay_\mu(s_{13},s_{23})_s\equiv \see_1(s, s_{13},s_{23})p_{1\mu}-\see_2(s, s_{13},s_{23})p_{2\mu} \,.
\label{eq:17_sn}
\ee
These form factors could be made explicit in a quasi-two{-}body approach to the $K_1$ decay\cite{nous1, note-profile}. Here, on the contrary, we want to determine them in a model independent way by using the experimental data to avoid the ambiguities described in the introduction. 

%%%%%%%%%%%%%%%%%%%%%%%%%%%%%%%%%%%%%%%%%%
\section{Angular distribution}
Now, we define the probability density function (PDF) for {a} given value of $s$. 
First, the different transverse ($s_z=\pm$) and the longitudinal ($s_z=0$) polarizations of $V$ state do not interfere, thus the decay rate is written as\footnote{{For $V=J/\psi$, we integrate over the $J/\psi$  decay angle here so that the interference term disappears.}} :
%\bea 
%\frac{d\Gamma(B\to K_{1} V\to (K\pi\pi)V)_s}{ds_{13}ds_{23}d(\cos\theta)d\phi} 
%&=& 
%\frac{(2\pi)^4}{2M_B}(2\pi)^3%\sqrt{s}
%\emi{ds}\frac{1}{(2\pi)^5} \frac{|\vec{p}_V^{{\,*}}|}{\emi{2}M_B}
% \no
%&& \times
%%%\frac{(2\pi)^4}{2\sqrt{s}}
%\frac{1}{32(2\pi)^8s}
%%%\Gamma(\Bbar\to \Kbar_{1}(1270)_{s_z}\gamma_{s_z})
%%% ds_{12} ds_{13} d(\cos\theta)d\phi 
% \left|\frac{1}{(s-m_{K_1}^2)+im_{K_1} \Gamma_{K_1}(s)}\right|^2 
%\no
% && \times \sum_{s_z} |{\mathcal{A}_{s_z}^V}(s)|^2 \left|\vec{\epsilon}_{K_{1s_z}} \cdot \vec{\jay}_{K_1}(s_{13},s_{23})_s\right|^2 
%\label{eq:6_v1}
%\eea
\begin{equation}
\begin{split}
\frac{d\Gamma(B\to K_{1} V\to (K\pi\pi)V)_s}{ds_{13}ds_{23}d(\cos\theta)d\phi} =& \frac{(2\pi)^4}{2M_B}(2\pi)^3
{ds}\frac{1}{(2\pi)^5} \frac{|\vec{p}_V^{{\,*}}|}{{2}M_B} \\
& \times
\frac{1}{32(2\pi)^8s}
 \left|\frac{1}{(s-m_{K_1}^2)+im_{K_1} \Gamma_{K_1}(s)}\right|^2 \\
 & \times \sum_{s_z} |{\mathcal{A}_{s_z}^V}(s)|^2 \left|\vec{\epsilon}_{K_{1s_z}} \cdot \vec{\jay}_{K_1}(s_{13},s_{23})_s\right|^2 \,,\label{eq:6_v1}
\end{split}
\end{equation}
{where $\vec{p}_V^{{\,*}}$ is the three momentum of $V$ in the $B$ reference frame, while the $K_1$ polarization vector $\vec{\epsilon}_{K_1}$ and $\vec{\jay}_{K_1}$ are defined in the $K_1$ reference frame.}
%where the factor $\tilde{N} ^V_s$ is introduced in order to normalize the PDF for a given value of $s$: 
Note that in Eq.~\eqref{eq:6_v1}, the {\it width} in the denominator could also be related to $\vec{\jay}_{K_1}$, except, {we have to add all charge combinations, $K^+_1\to K^+\pi^+\pi^-$ and $K^+_1\to K^0 \pi^+\pi^0$ for $K^+_1$ and $K^0_1\to K^+\pi^0\pi^-$ and $K^0_1\to K^0 \pi^+\pi^-$ for $K^0_1$ (and similar for the charge conjugations).} 

{The PDF $\mathcal{W}^V(s_{13}, s_{23}, \cos\theta, \phi)_s$ is obtained from Eq.~\eqref{eq:6_v1} and is normalized as :}
\be \label{norm}
  \int ds_{13}\int ds_{23}\int d(\cos\theta) \int d\phi  \ \mathcal{W}^V(s_{13}, s_{23}, \cos\theta, \phi)_s =1 \,.
\ee
{Thus, the PDF can be written in terms of the squared decay amplitudes, which are the function{s} of the kinematical variables we are interested in, without the irrelevant pre-factors : }
%The factor $\tilde{N} ^V_s$ is a function of $s$, which cancels all the pre-factors in 
%Eq.~(\ref{eq:6_v1}) in front of the sum over $s_z$ to obtain the normalisation of 
%$\mathcal{W}^V(s_{13}, s_{23}, \cos\theta, \phi)_s$, thus, 
\be \label{normal}
\mathcal{W}^V(s_{13}, s_{23}, \cos\theta, \phi)_s=
\frac {\sum_{s_z}|{\mathcal{A}_{s_z}^V}(s)|^2 \left|\vec{\epsilon}_{K_{1s_z}} \cdot \vec{\jay}_{K_1}(s_{13},s_{23})_s\right|^2}
{\int ds_{13}\int ds_{23}\int d(\cos\theta)\int d\phi \sum_{s_z} |{\mathcal{A}_{s_z}^V}(s)|^2\left|\vec{\epsilon}_{K_{1s_z}} \cdot \vec{\jay}_{K_1}(s_{13},s_{23})_s\right|^2}
\ee

Next we make explicit the angular distribution of $\mathcal{W}^V$ (the definition of the coordinate system and angles is given in the Appendix) : 
\begin{equation}
{\mathcal{W}}^V(s_{13}, s_{23}, \cos\theta, \phi)_s\equiv  a^V+(a_1^V+a_2^V\cos 2\phi+a_3^V\sin 2\phi)\sin^{{2}}\theta+b^V \cos \theta \,,
\end{equation}
where the angular coefficients depend on the Dalitz variables {and fixed value of $s$}. { They can be written as :}
\begin{eqnarray}
a^V(s, s_{13}, s_{23})&=&{{N} ^V_s}\xi_a^V\left[|c_1|^2+|c_2|^2-2{\rm Re}(c_1c_2^*)\cos\delta \right] \label{Eq:4_sn} \,, \\
a^V_1(s, s_{13}, s_{23})&=&{N^V_s}\xi_{a_i}^V\left[|c_1|^2+|c_2|^2-2{\rm Re}(c_1c_2^*)\cos\delta \right] \label{Eq:5_sn} \,, \\
a^V_2(s, s_{13}, s_{23})&=&{N ^V_s}\xi_{a_i}^V\left[(|c_1|^2+|c_2|^2)\cos\delta -2{\rm Re}(c_1c_2^*) \right] \label{Eq:6_sn} \,, \\
a^V_3(s, s_{13}, s_{23}) &= &{N ^V_s}\xi_{a_i}^V\left[(|c_1|^2-|c_2|^2)\sin \delta \right] \label{Eq:7_sn} \,, \\
b^V(s, s_{13}, s_{23})& =&{-}{N ^V_s}\xi_{b}^V\left[2{\rm Im}(c_1c_2^*)\sin\delta \right] \,,
\label{Eq:8_sn}
\end{eqnarray}
{where the factor  $N^V_s>0$ is the normalization factor, which is equal to the inverse of the denominator of Eq.~\eqref{normal}}. 
%includes the normalization factor $\tilde{N}^V_s$ together with the pre-factor in Eq.~(\ref{eq:6_v1}) in front $\sum_{s_z}$ so that we are now left only with the ratios of the angular coefficients}.  

The $\xi$'s represent the $B\to K_1 V$ decay, {and} thus, depend only on $s$
%\bea
%\xi_a^V(s)&\equiv& \frac{|{\mathcal{A}}_+^V(s)|^2+|{\mathcal{A}}_-^V(s)|^2}{2}, \no
%\xi_{a_i}^V(s)&\equiv& \frac{-(|{\mathcal{A}}_+^V(s)|^2+|{\mathcal{A}}_-^V(s)|^2)+2|{\mathcal{A}}_0^V(s)|^2}{4}, \no
%\xi_{b}^V(s)&\equiv& \frac{|{\mathcal{A}}_+^V(s)|^2-|{\mathcal{A}}_-^V(s)|^2}{2} \,.
%\eea
\begin{equation}
\begin{split}
\xi_a^V(s) \equiv& \frac{|{\mathcal{A}}_+^V(s)|^2+|{\mathcal{A}}_-^V(s)|^2}{2} \,, \\
\xi_{a_i}^V(s) \equiv& \frac{-(|{\mathcal{A}}_+^V(s)|^2+|{\mathcal{A}}_-^V(s)|^2)+2|{\mathcal{A}}_0^V(s)|^2}{4} \,, \\
\xi_{b}^V(s) \equiv& \frac{|{\mathcal{A}}_+^V(s)|^2-|{\mathcal{A}}_-^V(s)|^2}{2} \,.
\end{split}
\label{eq:xis}
\end{equation}
In fact, for $V=\gamma$, {the longitudinal amplitude vanishes ($\mathcal{A}_0^\gamma=0$), {which} simplifies the above expressions, giving as a result $a^\gamma=-2a_1^\gamma$.}

The coefficients $c_{1,2}$ are related to the form factors in Eq.~\eqref{eq:17_sn} as : 
\[c_1(s, s_{13}, s_{23})={\mathcal{C}_1(s, s_{13}, s_{23})}|\vec{p}_1|, \quad c_2(s, s_{13}, s_{23})={\mathcal{C}_2(s, s_{13}, s_{23})}|\vec{p}_2| \,, \]
where we wrote explicitly the Dalitz variable{s} dependence. The angle $\delta$ (with $0<\delta< \pi $) is defined as
\[\cos\delta = \frac{\vec{p}_1\cdot \vec{p}_2}{|\vec{p}_1||\vec{p}_2|} \,. \]
Let us also remind that all the relevant kinematical variables  can be expressed in terms of the Dalitz variables : 
\[|\vec{p}_{1,2}|^2=E_{1,2}^2-m_{1,2}^2 \,, \quad \vec{p}_1\cdot \vec{p}_2=E_1E_2-\frac{s_{12}-m_1^2-m_2^2}{2} \,, \quad E_{1,2}=\frac{s-s_{23, 13}+m_{1,2}^2}{2\sqrt{s}} \,. \]

%%%%%%%%%%%%%%%%%%%%%%%%%%%%%%%%
%%%%%%%%%%%%%%%%%%%%%%%%%%%%%%%%
\section{Photon polarization : relating the $\bm{B\to K_1\gamma}$ and $\bm{B\to K_1 J/\psi}$ amplitudes}
The photon polarization {in} the $B\to K_1\gamma$ process which we want to determine is {defined as following :}
\begin{equation}
\lambda_\gamma \equiv \frac{|{\mathcal{A}}_+^\gamma(s)|^2-|{\mathcal{A}}_-^\gamma(s)|^2}{|{\mathcal{A}}_+^\gamma(s)|^2+|{\mathcal{A}}_-^\gamma(s)|^2} \,,
\end{equation}
{where in the SM, $\lambda_\gamma \simeq +1 (-1)$ for $B^{{0}}, B^+ (\Bbar^{{0}}, B^-)$.}
{In this article, we do not discuss the so-called charm loop contributions, which may differentiate slightly $\lambda_\gamma$ from $\pm 1$. Under this assumption,}
%the \andrey{$B\to K_1$} form factor\andrey{s} for $\mathcal{A}^\gamma_+$ and $\mathcal{A}^\gamma_-$ are the same, which cancel all the $s$ dependence, i.e. $\xi^\gamma_{a, a_{i}, b}(s)=\xi^\gamma_{a, a_{i}, b}$.}
{the $s$-dependence of $\mathcal{A}_\pm^\gamma(s) \propto T_1(s)$, where $T_1$ is the $B\to K_1$ hadronic form factor, is cancelled out in the ratios. Hence,  one can write $\xi^\gamma_{a, a_{i}, b}(s)=\xi^\gamma_{a, a_{i}, b}$}\footnote{For the same reason, strictly speaking, $\lambda_\gamma$ here is slightly different from the usual definition of $\lambda_\gamma\equiv \frac{|C_+|^2-|C_-|^2}{|C_+|^2+|C_-|^2}$ where $C_\pm$ represents only the short-distance $b\to s\gamma$ decay.}.
{Using Eq.~\eqref{eq:xis}, one can find}
%By using the equations derived in the previous section, we can find
\begin{equation}
{ \lambda_\gamma=%2~\frac{\xi_{b}^\gamma }{3\xi_a^\gamma+2\xi_{a_i}^\gamma} =
\frac{\xi_{b}^\gamma }{\xi_a^\gamma} } \,.
\label{eq:lambda_def}
\end{equation}

In the following, we show that the $\xi_{a, a_i, b}^\gamma $'s can be indeed obtained from the measurement of $a^V, a_i^V, a^\gamma, b^\gamma$ {\it in a model independent way}. 

First, we obtain  $\xi_{a}^\gamma$ via : 
\begin{equation}
\xi_a^\gamma=\frac{a^\gamma(s, s_{13}, s_{23})}{N_s^\gamma\left[|c_1|^2+|c_2|^2-2{\rm Re}(c_1c_2^*)\cos\delta\right]} \,.
\label{eq:xiagamma}
\end{equation}
The term in the square brackets in the denominator is common for $V=J/\psi, \gamma$ and can be obtained for given point of $(s, s_{13}, s_{23})$  as 
\begin{equation}
|c_1|^2+|c_2|^2-2{\rm Re}(c_1c_2^*)\cos\delta = \frac{a^V(s, s_{13}, s_{23})}{N ^V_s\xi_a^V(s)}= \frac{ a^V_1(s, s_{13}, s_{23})}{N ^V_s\xi_{a_i}^V(s)} \,.
\label{eq:15_vav}
\end{equation}

Next, we determine $\xi_{b}^\gamma$ from the experimental measurement of $b^\gamma(s, s_{13}, s_{23})$ : 
\begin{equation}
\xi_b^\gamma=-\frac{b^\gamma(s, s_{13}, s_{23})}{N_s^\gamma\left[2~{\rm Im}(c_1c_2^*)\sin\delta\right] } \,.
\label{eq:xibgamma}
\end{equation}
Now we obtain the denominator factor {$2{\rm Im}(c_1c_2^*)\sin\delta$}. By writing 
\[ {\rm Im}(c_1c_2^*)=\pm \sqrt{|c_1|^2|c_2|^2-[{\rm Re}(c_1c_2^*)]^2} \,, \] 
we find that we need to obtain independently these two factors, $|c_1|^2|c_2|^2$ {and} ${\rm Re}(c_1c_2^*)$, from the above equations. 
Then, by using {Eqs.~\eqref{Eq:5_sn}-\eqref{Eq:7_sn}}, we find 
\begin{equation}
2~{\rm Im}(c_1c_2^*)\sin\delta = \pm \frac{1}{N ^V_s\xi_{a_{i}}^V(s)} \sqrt{(a_1^V(s, s_{13}, s_{23}))^2-(a_2^V(s, s_{13}, s_{23}))^{2}-(a_3^V(s, s_{13}, s_{23}))^{2}}  \label{Eq:11_sn} 
\end{equation}

Finally, the sign ambiguity remains, which can not be resolved at this point.

%Now by inserting Eqs. (\eqref{eq:xiagamma}), (\ref{eq:15_vav}) (\ref{eq:xibgamma}) and (\ref{Eq:11_sn}) into Eq.~\eqref{eq:lambda_def}, % and using also the relation between $N^V$ and $N^{\gamma}$, eqn. (\ref{N}), 
Now by inserting {Eqs.~\eqref{eq:xiagamma}-\eqref{Eq:11_sn}} into Eq.~\eqref{eq:lambda_def}, we can obtain the polarization parameter which we want to determine :
%\bea
%\lambda_\gamma&=&\frac{\xi_{b}^\gamma } {\xi_{a}^{\gamma}}\label{Eq:16_sn}\\
%&=&\mp \frac {b^\gamma(s, s_{13}, s_{23})}{a^\gamma(s, s_{13}, s_{23}) } \times \frac {1}{\sqrt{1-\left(\frac{a_2^V(s, s_{13}, s_{23})}{a_1^V(s, s_{13}, s_{23})}\right)^2-\left(\frac{a_3^V(s, s_{13}, s_{23})}{a_1^V(s, s_{13}, s_{23})}\right)^2}} \nonumber
%\eea
{
\begin{equation}
\label{Eq:16_sn}
\lambda_\gamma = \frac{\xi_{b}^\gamma } {\xi_{a}^{\gamma}} = \mp \frac {b^\gamma(s, s_{13}, s_{23})}{a^\gamma(s, s_{13}, s_{23}) } \times \frac {1}{\sqrt{1-\left(\frac{a_2^V(s, s_{13}, s_{23})}{a_1^V(s, s_{13}, s_{23})}\right)^2-\left(\frac{a_3^V(s, s_{13}, s_{23})}{a_1^V(s, s_{13}, s_{23})}\right)^2}} \,.
\end{equation}
}
The right hand side of Eq.~\eqref{Eq:16_sn} is the main result of this paper. 
This equation implies :
\begin{itemize}
\item The photon polarization {in} $B\to K_1\gamma$ can be obtained from the measurement of the angular coefficients $a^\gamma(s, s_{13}, s_{23})$, $b^\gamma(s, s_{13}, s_{23})$ which can be measured only via the standard $\cos\theta$ distribution, together with  the coefficients $a^V_{1,2,3}(s, s_{13}, s_{23})$ which requires the azimuthal angle $\phi$ distribution. The advantage is that the latter coefficients can be measured equally by using either $B\to J/\psi K_1$ or $B\to K_1\gamma$ decays. Therefore, we can take advantage of the much higher statistics of the $J/\psi$ process. 
\item The {final} results depend only on the ratio of the angular coefficients so that there is no need for the normalization. 
\item The photon polarization $\lambda_\gamma$ does not depend on {$s$ nor any Dalitz variables} 
{(except for the neglected charm contribution mentioned in the section 2)}, which implies that the {expression in} Eq.~\eqref{Eq:16_sn} is  constant  at any point of the  $(s, s_{13}, s_{23})$ plane. When we use the $J/\psi$ to determine the denominator of this term, we simply  need to map point by point on the Dalitz plane. 
\item {Concerning the sign ambiguity, in practice, we may measure the absolute value of the polarization parameter $|\lambda_\gamma|$. In this way, we are left with the sign ambiguity of overall sign of $\lambda_\gamma$ but we can neglect the sign variation of $b^\gamma/a^\gamma$ term since $\lambda_\gamma$ must be constant in the $(s, s_{13}, s_{23})$ plane. }
\end{itemize}
The third point has important consequence: arbitrary binning may lead to a variation of $\lambda_\gamma$ depending on the Dalitz points. Having the large sample available in $B\to K_1 J/\psi$  ({$\sim\mathcal{O}(10^3)$ events in the $K_1(1270)$ region even at Belle~\cite{Guler:2010if}, which means orders of magnitudes higher at LHCb)}, a high sensitivity to $\lambda_\gamma$ is expected. 
 Nevertheless, the reliability of method has to be confirmed with a Monte Carlo simulation. In particular, the optimization of the binning could be used by modeling the resonances in a crude manner.

%%%%%%%%%%%%%%%%%%%%%%%%%%%%%%%%
\section{Conclusions}
The angular distribution in the polar angle $\theta$ of the 
{$B\to K_{\rm res}\gamma \to K\pi\pi\gamma$} process has recently been measured by the LHCb collaboration~\cite{Aaij:2014wgo}. {Among various kaonic resonances $K_{\rm res}$, a large $B\to K_1(1270)\gamma$ contribution has been identified, confirming the previous result~\cite{Yang:2004as}.}
The extraction of the $b\to s\gamma$ photon polarization from this data requires a detailed knowledge of the $K_1$ decays, in particular, the imaginary part of the product of the two form factors, ${\rm Im}(c_1c_2^*)$. The imaginary part is, in general, very sensitive to the resonance structure of the decay while there are many uncertainties in the resonance decay structure of $K_1(1270)$,  especially due to i) the limited phase space for the main decay channel $K_1(1270)\to \rho K$ resulting in strong distortion effects, ii) a possible $K_1(1270)\to \kappa \pi$ contributions, {neither} well determined experimentally {nor} theoretically tractable.

In order to circumvent this problem, we propose a resonance model independent determination of  the strong interaction factor ${\rm Im}(c_1c_2^*)$. This method requires the Datliz plot of the angular coefficients including both polar and azimuthal angles. In this article, we have shown that the same Dalitz plot {analysis} can be also obtained through the $B\to J/\psi K_1\to J/\psi K\pi\pi$ channel. The $B$ decay part of these two channels are very different while we found that we have enough observables to separate the $B$ decay part. The realization of our proposal would require a detailed Monte Carlo studies, in particular by evaluating the binning effect.

\section*{Acknowledgements} 
%\label{sec:6}
We would like to thank Fran\c{c}ois Le Diberder for many discussions, in particular, on the feasibility of the method. 
We also aknowledge Patrick Roudeau, Akimasa Ishikawa and Yoshimasa Ono for discussions. 
{The work of A.T. has been carried out thanks to the support of the OCEVU Labex (ANR-11-LABX-0060) and the A*MIDEX project (ANR-11-IDEX-0001-02) funded by the "Investissements d'Avenir" French government program managed by the ANR.}

%%%%%%%%%%%%%%%%%%%%%%%%%%%%%%%%%%
%
\appendix
\section{Kinematics of $\bm{B^+\to V K_1^+\to VK^+\pi^+\pi^-}$ decay ($\bm{V=J/\psi, \gamma}$)}\label{definitions} 
In this section, we describe all the definitions of the kinematical variables. {We use  $B^+\to V K_1^+\to VK^+\pi^+\pi^-$ decay as an example but one can obtain the similar formulae for other charge combinations.} 
Throughout this article, we work {in} the $K_1$ rest frame. We can move to the conventional $B$  rest frame or any other frame simply {by} a Lorentz transformation. 
First, we assign the three moment{a} as 
\begin{equation}
\pi^+(\vec{p}_1) \,, \quad \pi^-(\vec{p}_2) \,, \quad K^+(\vec{p}_3) \,.
\end{equation}

%\begin{figure}[t]
%\begin{center}
%\epsfig{file=decayplot.eps,width=0.5\linewidth}
%\caption{The $K_1\to K\pi\pi$ decay plane in the  rest frame of $K_1$. 
%Defining the $-z$ direction as the photon direction, the $\theta$ is given as 
%$\cos\theta\equiv\left(\frac{\vec{p}_{1}\times \vec{p}_{2}}{|\vec{p}_{1} \times \vec{p}_{2}|}\right)_z$.}
%\label{fig:1}
%\end{center}
%\end{figure}

Now, we define a standard orthogonal frame, with respect to the spin direction of $K_1$, or ${V=}J/\psi,\gamma$. First, the $Oz$ is defined as the $V$ direction
\begin{equation}
\vec{e}_z=\frac{\vec{p}_V}{|\vec{p}_V|}=\frac{-\vec{p}_B}{|\vec{p}_B|} \,.
\end{equation}
We define the axis perpendicular to the $K\pi\pi$ decay plane by $\vec{n}$ : 
\begin{equation}
%\vec{n}=\frac{\vec{p_{\pi^+}}\times \vec{p_{\pi^-}}}{|\vec{p_{\pi^+}}. \vec{p_{\pi^-}}|}
{\vec{n}=\frac{\vec{p}_1\times \vec{p}_2}{|\vec{p}_1 {\times} \vec{p}_2|}} \,.
\end{equation}
Then, the $Oy$ is chosen as normal to the $Oz$ and $V=J/\psi, \gamma$ direction by 
\begin{eqnarray}
\vec{e}_y=\frac { \vec{p}_V \times \vec{n}} {|\vec{p}_V {\times} \vec{n}|} \,.
\end{eqnarray}
Finally, $Ox$ is then chosen as the normal to $Oy$ {and} $Oz$ : $ \vec{e}_x=\vec{e}_y \times \vec{e}_z$ .

One also defines a polar angle $\theta$, of $\vec{n}$ with respect to the $\vec{e}_z$ :
\be
\cos\theta= \vec{e}_z \cdot \vec{n} \label{costheta}
\ee
 Let us here set a condition for $\theta$ as 
\be
\vec{e}_x\cdot \vec{n}=\sin\theta >0, \quad 0<\theta<\pi \,.
\ee

\begin{figure}[t!]\centering
\includegraphics[height=0.5\textwidth]{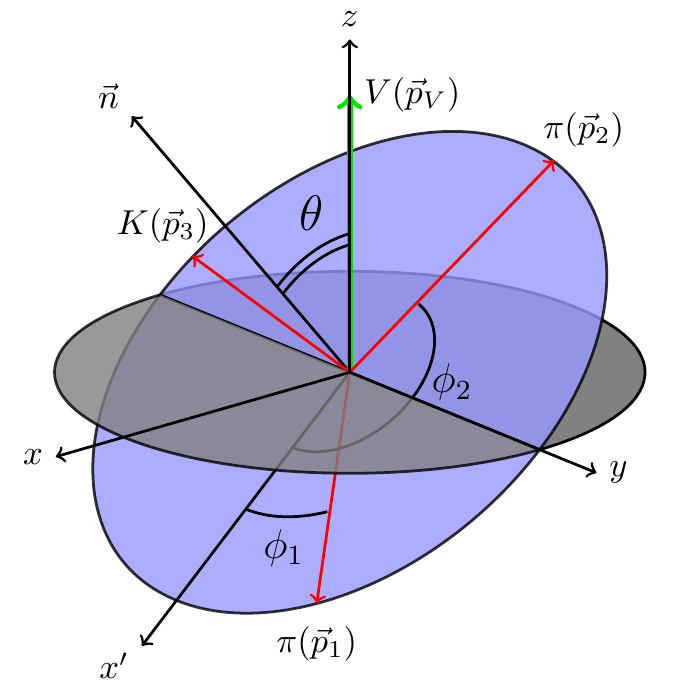}
\caption{\footnotesize Kinematics of the $B\to K_1(\to K\pi\pi)V$ decay.}
\label{fig:angles}
\end{figure}

Now we rotate  $\vec{e}_x$ onto the {$K\pi\pi$} decay plane and define the result as $\vec{e}_x^{\,\prime}$ which can be written as 
\be
\vec{e}_x^{\,\prime} = \vec{e}_y \times \vec{n}
\ee 
We can then define a second orthogonal frame, which is based on the $K_1$ decay plane, $\vec{e}^{\,\prime},\vec{e}_y, \vec{n}$.  Defining $\phi_{1,2}$ to be the azimuthal angle from the $\vec{e}_x^{\,\prime}$ axis in this $(x^\prime,y)$ decay plane, {the components of the pions three momenta,}
\be
\vec{p}_{1,2} = {|\vec{p}_{1,2}|} (\cos \phi_{1,2}~\vec{e}_x^{\,\prime} + \sin \phi_{1,2}~\vec{e}_y ) \,,
\label{eq:p12}
\ee
{can be expressed in terms of $\theta,\phi_{1,2}$ in the standard frame as :}
%, by making the scalar products with $\vec{e}_{x,y,z}$ :
%\begin{eqnarray}
%\label{eq:8}
%&(\vec{p}_{1})_x&=|\vec{p}_1| \cos \theta \cos \phi_1 \\ \nonumber
%&(\vec{p}_{1})_y&=|\vec{p}_1| \sin \phi_1 \\ \nonumber
%&(\vec{p}_{1})_z&=-|\vec{p}_1| \sin \theta \cos \phi_1 \\ \nonumber
%    {   }    \\ \nonumber
%&(\vec{p}_{2})_x&=|\vec{p}_2| \cos \theta \cos \phi_2 \\ \nonumber
%&(\vec{p}_{2})_y&=|\vec{p}_2| \sin \phi_2  \\ \nonumber
%&(\vec{p}_{2})_z&=-|\vec{p}_2| \sin \theta \cos \phi_2 \label{pi},
%\end{eqnarray}
\begin{equation}
\label{eq:p12components}
\begin{split}
(\vec{p}_{1,2})_x &=|\vec{p}_{1,2}| \cos \theta \cos \phi_{1,2} \,, \\
(\vec{p}_{1,2})_y &=|\vec{p}_{1,2}| \sin \phi_{1,2} \,, \\
(\vec{p}_{1,2})_z &=-|\vec{p}_{1,2}| \sin \theta \cos \phi_{1,2} \,.
\end{split}
\end{equation}

The advantage is that the angles $\theta,\phi_{1,2}$ are connected directly with the decay plane. We note that the linear combination of the {$\phi_{1,2}$ angles}, 
\be
\delta\equiv \phi_2-\phi_1 \,,
\label{eq:delta}
\ee
is a function the Dalitz variables defined by
%\begin{eqnarray}
%s&=&(p_{K_1})^2 \no
%s_{13}&=&(p_1+p_3)^2=(p_{K_1}-p_2)^2 \\ \nonumber
%s_{23}&=&(p_2+p_3)^2=(p_{K_1}-p_1)^2\\ \nonumber
%s_{12}&=&(p_1+p_2)^2=(p_{K_1}-p_3)^2
%\end{eqnarray}
\begin{equation}
\begin{split}
s =& (p_{K_1})^2 \\
s_{13}=&(p_1+p_3)^2=(p_{K_1}-p_2)^2 \,, \\
s_{23}=&(p_2+p_3)^2=(p_{K_1}-p_1)^2 \,, \\
s_{12}=&(p_1+p_2)^2=(p_{K_1}-p_3)^2 \,.
\end{split}
\end{equation}
{In the $K_1$ rest frame, $\vec{p}_{K_1}=0$ and {$|\vec{p}_{1,2,3}|$ can be expressed in terms of $s_{23},s_{13},s_{12}$ respectively. Since only two of them are independent, we choose $s_{23},s_{13}$ for symmetry. Then the relative angle between the three momenta of the two pions}
\begin{eqnarray}
\cos\delta=\frac{\vec{p}_1 \cdot \vec{p}_2}{|\vec{p}_1||\vec{p}_2|}= \frac {|\vec{p}_3|^2-|\vec{p}_1|^2-|\vec{p}_2|^2} {2|\vec{p}_1||\vec{p}_2|} \,,
\label{delta}
\end{eqnarray}}
is expressible in terms of {$s, s_{13},s_{23}$}. {The same holds} for the other relative angles between the three momenta\footnote{We have furthermore $0<\delta<\pi,{({\sin\delta >0})}$, because the angles $\phi_1,\phi_2$ are measured in the plane oriented by the normal $\vec{n}=\vec{p}_1\times\vec{p}_2{/|\vec{p}_1\times\vec{p}_2|}$.}. This means that the $K \pi \pi$ system is rigid once the masses of the two $K \pi$ subsystems have been chosen. It is still allowed to rotate however : if the normal is fixed by a definite $\theta$, there remains a free rotation of the rigid $K\pi\pi$ system around $\vec{n}$ in the decay plane. We choose the angle defining this rotation as :
\be \label{def.phi}
\phi \equiv \frac{\phi_1+\phi_2}{2} \,. 
\ee
In this way, the angle $\phi$ {in the reference \cite{Gronau:2001ng} is now {fixed, which  allows to perform definite calculations}}. Note that our definition is just one possible among many others {while we have found it} convenient because {it simplifies the} calculations.

{
Then, {re-expressing} $\phi_{1,2}$ as
%\begin{eqnarray} \label{phi}
%\phi_1=\phi-\frac {\delta}{2}\\ \nonumber
%\phi_2=\phi+\frac {\delta}{2}
%\end{eqnarray}
\begin{equation}
\label{phi}
\phi_{1,2} = \phi \mp \frac {\delta}{2} \,, \nonumber
\end{equation}
one can get the components of $\vec{p}_{1,2}$ in Eq.~\eqref{eq:p12components}, expressed in terms of $\phi$ and the Dalitz variables.
}

%%%%%%%%%%%%%%%%%%%%%%%%%%%%%%%%%%%%%%%%%%%%%%%%%%% 

\end{document}